# Theory of fishnet negative-index optical metamaterials

J. Yang[1], C. Sauvan[1], H.T. Liu[2] and P. Lalanne[1,%*]

[1]Laboratoire Charles Fabry de l'Institut d'Optique, CNRS, Univ Paris-Sud, Campus Polytechnique, RD 128, F-91127 Palaiseau, France

[2]Key Laboratory of Opto-electronic Information Science and Technology, Ministry of Education, Institute of Modern Optics, Nankai University, Tianjin 300071, China.

[%]also at Laboratoire Photonique Numérique et Nanosciences, Université Bordeaux 1, CNRS, Institut d'Optique, F-33405 Talence, France.

[*]e-mail: philippe.lalanne@institutoptique.fr

## Abstract

We theoretically study fishnet metamaterials at optical frequencies. In contrast to earlier works, we provide a microscopic description by tracking the transversal and longitudinal flows of energy through the fishnet mesh composed of intersecting subwavelength plasmonic waveguides. The analysis is supported by a semi-analytical model based on surface-plasmon coupled-mode equations, which provides accurate formulas for the fishnet refractive index, including the real-negative and imaginary parts. The model simply explains how the surface plasmons couple at the waveguide intersections and it shines new light on the fishnet negative-index paradigm at optical frequencies. Extension of the theory for loss-compensated metamaterials with gain media is also presented.

In the recently emerged fields of metamaterials and transformation optics, the possibility of creating optical negative-index metamaterials (NIMs) using nanostructured metal-dielectric composites has triggered intense basic and applied research over the past several years [1-4]. In view of potential applications of such structures in a variety of areas ranging from subwavelength imaging [1] to cloaking [5], it is important to understand the underlying physics in detail. Until now, our common understanding of optical NIMs is entirely based on the concept of homogenisation. Different techniques have been developed, including field averaging [6], Bloch mode computation [7-9], multipole expansion [10,11], and inversion of scattering parameters [12], all of them being based on fully-vectorial electromagnetic calculations of the whole structure. As a consequence, even if these approaches quantitatively predict the light transport, they are often non-intuitive, thereby hindering the design process required to apply metamaterials into new optical technologies. At microwave frequencies, lumped-elements circuit models provide a reliable framework for understanding and designing the properties of NIMs [13-14]. Unfortunately, such a framework is lacking in the optical domain.

In this Letter, we study fishnet NIMs [3-4,7-9] by adopting a "microscopic" point of view. We abandon classical homogenisation approaches and instead we track the energy as it propagates and scatters through the fishnet mesh, like a fluid flowing in a multi-channel system. The dynamics involves two intersecting subwavelength channels and their coupling, see Fig. 1. The longitudinal ($z$-direction) channel consists of air holes in a metal film and the transversal ($x$-direction) channel is formed by metal-insulator-metal (MIM) waveguides that support the propagation of surface plasmon polariton modes called gap-SPPs. Through simple coupled-mode equations that describe how both channels exchange energy, we derive semianalytical formulas for the refractive index and the losses, and we explain how low-loss and negative values appear. The model can be seen as an extension of the lumped-elements models used in the microwaves; the extension encompasses 1/ the treatment of ohmic losses that limit the operation at high frequency and 2/ the abandonment of periodic conditions [14] in the transversal direction. The latter allows for an analytical treatment of the key parameters impacting the transversal gap-SPP resonance, which shines new light on the physical origin of negative refraction in fishnets. In particular, we evidence that the reduction of the gap-SPP damping constitutes the main ingredient for the problem of loss-compensation with gain media, a vital issue for NIMs in the visible and near-infrared [2,15-17].

It is worth emphasizing that all the salient properties of a fishnet are driven by its fundamental Bloch mode. Indeed, the negative refraction deduced from prism-deviation measurements can be explained by considering Snell's law at an interface between air and a fishnet with a refractive index equal to the effective index $n_{eff}$ of the fundamental Bloch mode [4]. This observation has been recently confirmed by numerical results showing that the reflectance and the transmittance of finite-thickness fishnet slabs can be perfectly calculated by assuming that the energy transport inside the fishnet is solely due to the fundamental Bloch mode [9]. Therefore, the following analysis will be mainly devoted to understanding the origin of the negative $n_{eff}$ values. The latter have been calculated with a Fourier Bloch-mode method [9] at normal incidence for the geometrical parameters used in [4] and given in the caption of Fig. 1. The simulation results (circles in Fig. 2) predict a negative-index band for $\lambda > 1.7$ μm, with low loss around 1.8 μm.

In order to understand these important features, we model the energy transport in the fishnet mesh as resulting from the flow of surface plasmons through two intersecting subwavelength channels. Both channels, along with the definition of the associated scattering coefficients, are depicted in Fig. 1. The longitudinal channel consists of a one-dimensional (1D) hole chain in a metal film; it supports the propagation of the super-mode formed by the in-phase superposition of the $TE_{01}$ modes of every hole. At the hole chain boundaries, the super-mode is partly transmitted (coefficient $\tau$), reflected (coefficient $\rho$) or scattered (coefficient $\alpha$) into the gap-SPP mode of the transversal MIM channel, see Fig. 1(b). Similarly an incident gap-SPP is scattered by the hole chain, see Fig. 1(c). Because of Lorentz reciprocity theorem, the transversal scattering process only defines two additional coefficients denoted by $r_{sp}$ and $t_{sp}$. Starting from the sole knowledge of the scattering coefficients $\rho$, $\tau$, $\alpha$, $r_{sp}$ and $t_{sp}$ (calculated with a fully-vectorial modal method [19]), we now provide a step-by-step analysis that brings us to derive an analytical expression for $n_{eff}$.

We start by considering the most elementary building-block of the fishnet, the hole chain depicted in the inset (i1) of Fig. 2. The effective index $n_h$ of its fundamental super-mode is shown with the dashed-dotted black curves. The mode is evanescent at long wavelengths; $Im(n_h)$ rapidly increases as the wavelength exceeds the cut-off value of 1.4 µm, while $Re(n_h)$ remains nearly null. Note that the spectral range of evanescence coincides with that of negative index [14,20].

We next consider a more involved structure: a single hole chain etched into a Ag-MgF$_2$-Ag periodic stack, see Fig. 2(i2). The Bloch mode of the $z$-periodic hole-chain can be analytically derived by assuming that the field in the metallic holes is only formed by the superposition of two counter-propagating super-modes, a legitimate assumption for tiny holes. The transfer matrix that links the amplitudes of the forward and backward super-modes, the $A_m$'s and $B_m$'s in Fig. 2(i2), is a 2×2 matrix, $A_{m+1} = \tau v A_m + \rho v B_{m+1}$ and $B_m = \rho v A_m + \tau v B_{m+1}$, where $v = \exp(ik_0 n_h a_z)$, $a_z$ is the longitudinal period and $k_0 = 2\pi/\lambda$. The Bloch mode effective index $n_{pc}$ is easily derived from the matrix eigenvalues [21] and one gets

$$\cos(k_0\, n_{pc}\, a_z) = \frac{\tau^2 v^2 - \rho^2 v^2 + 1}{2\tau\, v}. \tag{1}$$

The longitudinal periodic structuring profoundly affects the nature of the energy transport [22], and, for wavelengths larger than the cut-off of the TE$_{01}$ mode, $n_{pc}$ is negative (dashed blue curves), an effect that we attribute to the launching of gap-SPPs into the transversal MIM waveguides. However, since $|Re(n_{pc})| << |Re(n_{eff})|$ and $Im(n_{pc}) >> Im(n_{eff})$, the longitudinal structuring cannot alone explain the low-loss negative index of the fishnet. This reveals the importance of another effect, namely a transversal resonant coupling that strengthens the gap-SPP excitation and that is responsible for the appearance of large negative $n_{eff}$ values.

We finally consider the whole fishnet structure, which can be viewed as an array of z-periodic hole-chains that interact through the excitation of gap-SPPs. A closed-form expression for the fishnet effective index $n_{eff}$ can be analytically derived by assuming that the energy transfer in the dielectric gaps is solely mediated by the fundamental gap-SPP mode with a symmetric magnetic field $H_y(-z) = H_y(z)$. Using the scattering coefficients $r_{sp}$, $t_{sp}$ and $\alpha$, we easily derive additional coupled-mode equations for the gap-SPP amplitudes $C_n$ and $D_n$ defined in Fig. 2(i3). With the periodicity condition along x, we eliminate the $C_n$'s and $D_n$'s to obtain a new 2×2 matrix, where the scattering coefficients $\rho$ and $\tau$ have been replaced by $\rho + \gamma$ and $\tau + \gamma$ [21]. After diagonalization, we obtain the fishnet effective index $n_{eff}$

$$\cos(k_0\, n_{eff}\, a_z) = \frac{(\tau + \gamma)^2 v^2 - (\rho + \gamma)^2 v^2 + 1}{2(\tau + \gamma)v} \ , \tag{2}$$

with $\gamma = 2\alpha^2 u/[1 - (t_{sp} + r_{sp})u]$, where $u = \exp(ik_0 n_{sp} a_x)$ is the gap-SPP phase-delay over one period and $k_0 n_{sp}$ is the gap-SPP propagation constant. The transversal coupling between the hole chains is fully described by a single parameter, $\gamma$, which physically represents the multiple scattering of the gap-SPPs in the MIM layers. The model predictions for the fishnet effective index are shown with the solid red curves in Fig. 2. They are found to quantitatively capture all the major features of the calculated data (circles), such as the negative $n_{eff}$ for $\lambda > 1.7$ μm, the increase (in absolute value) of $Re(n_{eff})$ with the wavelength, and the low-loss band for $1.8 < \lambda < 2$ μm followed by a rapid increase of the loss at longer wavelengths. Even the band-gap for $1.45 < \lambda < 1.75$ μm is accurately predicted. We have checked that the model accuracy is mainly limited by our assumption that the energy transfer through the metallic holes is solely due to the $TE_{01}$ mode (single-mode approximation) [21].

In addition to providing closed-form expressions for the effective index, the main force of the model is to explicitly dissociate longitudinal and transversal contributions. The dispersion curve of a single $z$-periodic hole-chain, Eq. (1) and dashed blue curves in Fig. 2, only reflects the longitudinal contribution and shows a large attenuation with a negligible negative index. Thus the transversal coupling mediated by gap-SPPs [described by the multiple-scattering parameter $\gamma$ in Eq. (2)] clearly appears as the origin for the appearance of large negative $n_{eff}$ values with low loss. In our model, the reduction of the attenuation, $Im(n_{eff}) \ll Im(n_{pc})$, is understood as follows: the gap-SPPs, which leak away in the case of a single hole-chain, are recycled back into nearby chains in the fishnet. The increase (in absolute value) of $Re(n_{eff})$ is more delicate to analyse; it is due to a transversal resonance at $\lambda \approx 2$ μm that boosts the excitation of gap-SPPs and thus enhances the negativity that was only emerging in the hole-chain. Figure 3(a) shows the Lorentzian shape of $|\gamma|^2$, which corresponds to the resonance of a periodic MIM waveguide coupled to two 2D arrays of semi-infinite vertical holes, see the inset. This resonance reflects into an enhancement of the gap-SPP field in the fishnet, as shown by the red curve in Fig. 3(b), which shows the normalized gap-SPP intensity $|C_n/A_m|^2$ ($D_n = C_n$ for normal incidence). The model predicts that $|C_n/A_m|^2$ increases by a factor 10 as the wavelength varies from 1.7 μm to 2 μm, where a maximum is reached. This prediction is found to be in good agreement with fully-vectorial computations (circles) and quantitatively supports the usual understanding that attributes the "magnetic" response of fishnets to an anti-symmetric current distribution in the MIM stack [1,4,20,23]. Additionally, the model allows us to calculate the $1/e^2$ decay length of the gap-SPP resonance, which is found to be delocalised over 4 periods.

Reducing the attenuation of NIMs operating in the visible and near-infrared is crucial and the issue of loss compensation with gain media has recently received much attention [15-17]. Because of its analytical treatment, the model constitutes a powerful tool to study and design fishnets with embedded gain. We have stressed the major impact of the transversal resonance, both on the negative index and on the attenuation, and thus we start by considering the impact of gain on the resonance lifetime of $\gamma$. Since $|r_{sp} + t_{sp}| \approx 1$ in the denominator of $\gamma$, the quality factor $Q$ of the resonance is mainly limited by the gap-SPP damping, $Q \approx \mathrm{Re}(n_{sp})/[2\mathrm{Im}(n_{sp})]$. Numerical calculations have confirmed that all the scattering coefficients in Eq. (2) ($\alpha$, $r_{sp}$, $t_{sp}$, $\rho$, $\tau$) weakly depend on gain, and since $v$ is unaffected (we incorporate gain only in the dielectric MgF$_2$ layers), gain is mainly impacting the gap-SPP normalized propagation constant $n_{sp}$. Thus, according to the model, the intricate problem of loss-compensation in a fishnet metamaterial essentially reduces to a much simpler problem of loss-compensation in a MIM waveguide. Hereafter we restrict the study to the low-loss band, $1.75 < \lambda < 1.95$ µm, and we assume that the amplification process can be simply analysed by a phenomenological refractive index $n_d = 1.38 - ig$ with a constant negative imaginary part. Consistently with previous studies [15-17], we choose $0 < g < 0.02$. These values are small enough to ensure a linear operation for which self-consistent calculations are not strictly required [16]. As anticipated, due to the decrease of the gap-SPP damping, see Fig. 4(a), the transversal resonance is strengthened in the presence of gain, and the $Q$-factor rapidly increases with $g$, see Fig. 3(a). Note that gain negligibly influences $Re(n_{sp})$ but largely affects the characteristic length of the transversal resonance that extends over 30 cells for $g = 0.022$.

We next study the loss-compensated fishnet effective index, which is still given by Eq. (2) in the presence of small gain. As shown in Fig. 4(b), $Re(n_{\mathrm{eff}})$ is insensitive to the modest gain increase and *remains negative*. More importantly, the attenuation $Im(n_{\mathrm{eff}})$ is significantly lowered as $g$ increases, see Fig. 4(c). When the metallic loss of the gap-SPP is exactly compensated by amplification, i.e., for $Im(n_{sp}) \to 0$, the resonance linewidth of $|\gamma|^2$ becomes limited by the scattering losses. They can be partly compensated for larger gain because the gap-SPP becomes slightly amplified, see Fig. 4(a). For $g = 0.022$, the fishnet becomes an amplifying medium [$Im(n_{eff}) < 0$] for $1.75 < \lambda < 1.85$ µm, as shown by the black curve in Fig. 4(c). We have tested the predictions obtained for the fundamental Bloch mode of an infinite fishnet by analysing a real situation with a finite-thickness fishnet composed of 5 unit cells and illuminated from air at normal incidence. Figures 4(d) and 4(e) compare the specular transmission T and reflection R, obtained either by the Fabry-Perot model (solid curves) assuming that the energy transport through the fishnet slab is solely mediated by the fundamental Bloch mode with an effective index $n_{eff}(g)$ [9] or by fully-vectorial calculations (circles) obtained with the Rigorous-Coupled-Wave-Analysis. Except for a systematic spectral shift that has been removed and that is due to a slight offset of the model predictions for $Re(n_{eff})$ already seen in Figs. 4(b) or 2(a), the model is found to be highly accurate despite its simplicity.

In summary, light transport in fishnet NIMs has been analysed with a comprehensive and accurate model. In addition to providing an analytical treatment, the model shines new light on how a negative index is formed and how the inevitable losses associated to a magnetic-like resonance can be compensated by incorporating gain. We have shown that gain mainly affects the attenuation, leaving the negative real part of the refractive index unchanged. However, at the transparency threshold, the transversal SPP resonance becomes delocalised over a few tens of unit cells and fishnets with gain should not be considered as 3D metamaterials, but rather as 1D layered systems. The "microscopic" formalism that tracks the local transport of electromagnetic fields in the structure is especially suited for metamaterials composed of meshed channels, such as the fishnet. Perhaps with a weaker analyticity of the treatment, it could however be generalized to other geometries based on localized SPP resonance, such as paired-nanorod or split-ring metamaterials [1]. We hope that this formalism may be helpful not only to design NIMs, but also to engineer complex metallo-dielectric surfaces in general [24].

**Acknowledgements**

H.T. Liu acknowledges a Poste Rouge fellowship from CNRS. We thank Jean-Paul Hugonin for computational assistance.

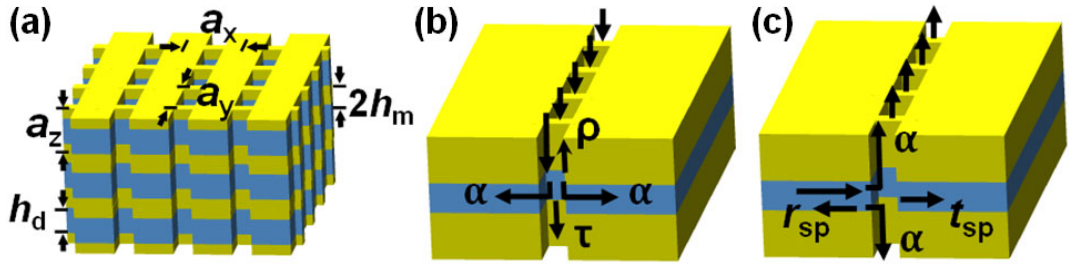

**Figure 1.** Elementary scattering events involved in a fishnet. **(a)** The fishnet under study is a 2D array (period $a_x = a_y = 860$ nm) of rectangular holes (width $w_x = 295$ nm and $w_y = 595$ nm) etched into a Ag($h_m = 50$ nm)-MgF$_2$($h_d = 50$ nm)-Ag($h_m = 50$ nm) periodic stack. The refractive index of MgF$_2$ is $n_d = 1.38$ and the frequency-dependent permittivity of silver is taken from [18]. **(b)** Scattering of the super-mode of a 1D hole chain. **(c)** Scattering of the gap-SPP mode supported by a MIM waveguide. The scattering events in **(b)** and **(c)** define five elementary scattering coefficients, the reflection $\rho$ and transmission $\tau$ of the super-mode, the reflection $r_{sp}$ and transmission $t_{sp}$ of the gap-SPP, and the coupling coefficient $\alpha$.

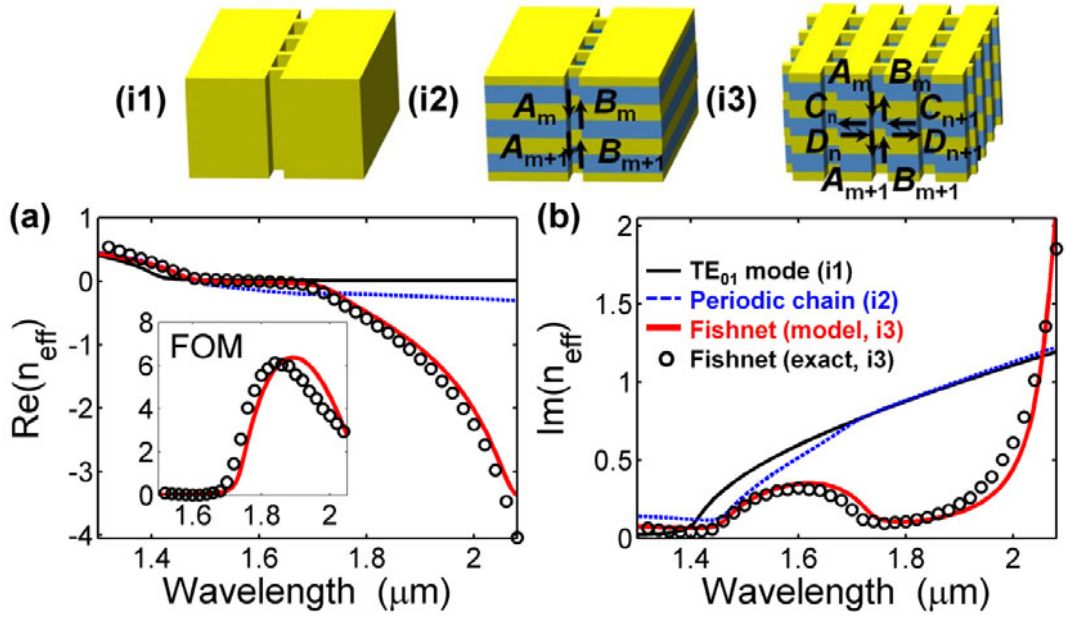

**Figure 2.** Model predictions for **(a)** the real part and **(b)** the imaginary part of the effective index. We study successively with the model the super-mode of a 1D hole chain [inset (i1), dashed-dotted black curves], the Bloch mode of a *single z*-periodic chain [inset (i2), dashed blue curves] and the Bloch mode of the fishnet [inset (i3), red curves]. Results of fully-vectorial calculations for the fishnet are marked with circles. The figure of merit (FOM) $|Re(n_{eff})|/Im(n_{eff})$ is shown in the inset of **(a)**.

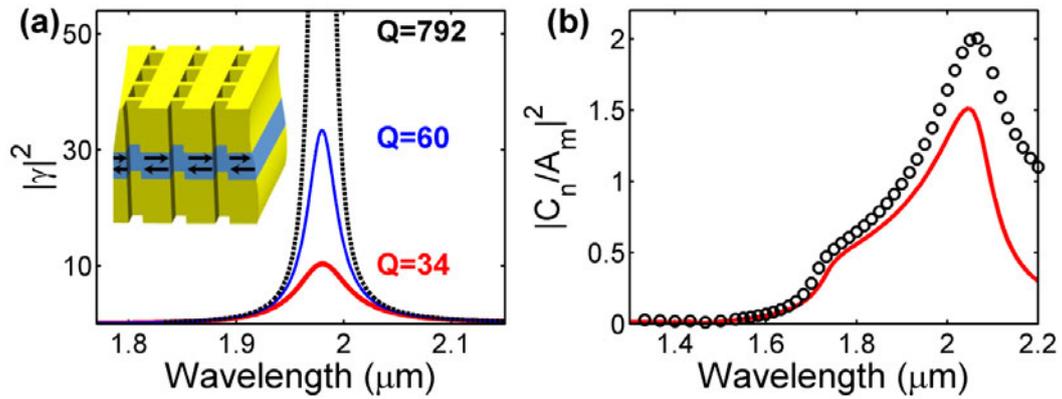

**Figure 3.** Transversal gap-SPP resonance. **(a)** Impact of the gain $g$ on the quality factor $Q$ of the resonance of $|\gamma|^2$, $g = 0$ (red), $g = 0.01$ (dashed blue) and $g = 0.022$ (dashed-dotted black). Inset: periodic MIM waveguide perforated by semi-infinite metallic holes. **(b)** Normalized gap-SPP intensity $|C_n/A_m|^2$ in the fishnet. Good agreement is achieved between the model predictions (red curve) and the values extracted from a fully-vectorial calculation of the fishnet Bloch mode (circles).

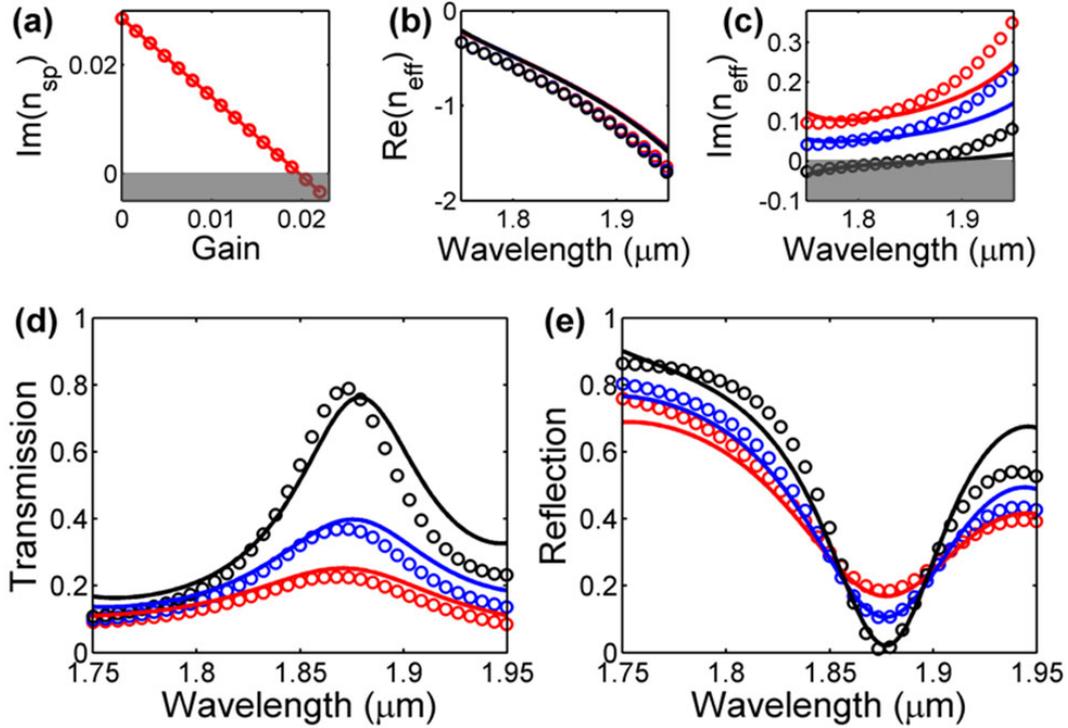

**Figure 4.** Loss compensation with gain. **(a)** Decrease of the gap-SPP damping with the gain $g$ ($\lambda = 1.9\ \mu m$). Amplifyied gap-SPPs are obtained in the grey region. **(b)** and **(c)** Fishnet $n_{eff}$ for three gain values, $g = 0$ (red), 0.01 (blue) and 0.022 (black). **(d)** and **(e)** Transmission and reflection spectra of a finite-thickness ($d = 5a_z$) fishnet for the same $g$'s values. In **(b)**, **(c)**, **(d)** and **(e)**, the model predictions are shown with solid curves and fully-vectorial calculations are shown with circles. For the sake of clarity, in **(d)** and **(e)**, the model predictions are blue-shifted by 20 nm to compensate for the slight offset in Re($n_{eff}$) due to the small metal thickness.

# Theory of fishnet negative-index optical metamaterials (EPAPS)

## J. Yang, C. Sauvan, H.T. Liu and P. Lalanne

We provide hereafter some technical elements concerning the analytical derivations and the numerical calculations presented in the main text.

## 1. Coupled-mode theory

The coupled-mode theory used to model the flow of light in a fishnet metamaterial relies on the single mode assumption. In the longitudinal ($z$-direction) channel, which consists of a one-dimensional (1D) hole chain in a metal film, we assume that the light transport is solely due to the propagation of an *evanescent super-mode* formed by the in-phase superposition of the fundamental $TE_{01}$ mode of every subwavelength hole. Similarly, in the transversal ($x$-direction) metal-insulator-metal (MIM) channels, the energy transport is assumed to be due to the fundamental gap-SPP mode of the MIM waveguide. The fundamental gap-SPP has a symmetric magnetic field profile, $H_y(-z) = H_y(z)$, and an antisymmetric electric field profile, $E_x(-z) = -E_x(z)$ [1].

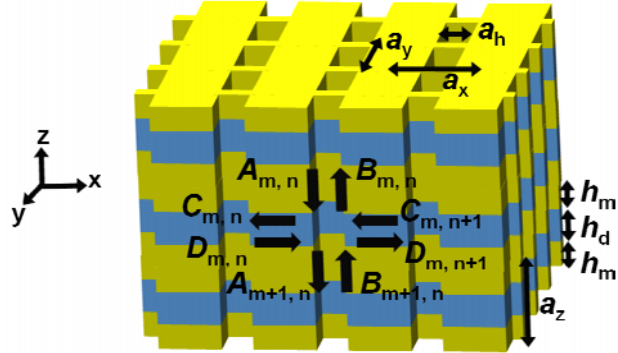

**Figure A.** Fishnet structure along with a definition of the modal amplitudes. In the longitudinal ($z$-direction) and transversal ($x$-direction) channels, light transport is only mediated by counter-propagating $TE_{01}$ super-modes (amplitudes $A_{m,n}$ and $B_{m,n}$) and gap-SPPs (amplitudes $C_{m,n}$ and $D_{m,n}$), respectively. The periods along $x$, $y$ and $z$-directions are denoted by $a_x$, $a_y$ and $a_z$. One fishnet unit cell consists of a metal (thickness $h_m$)-dielectric (thickness $h_d$)-metal (thickness $h_m$) stack with $a_z = 2h_m + h_d$.

As shown in Fig. 2(i3) in the main text, repeated here as Fig. A for the sake of clarity, we denote by $A_{m,n}$ and $B_{m,n}$ the amplitudes of the up-going and down-going $TE_{01}$ super-modes, and by $C_{m,n}$ and $D_{m,n}$ the amplitudes of the gap-SPP mode that propagates along the x-direction, leftward and rightward. The first subscript $m$ refers to the period number in the $z$-direction ($m^{th}$ metallic layer and $m^{th}$ dielectric layer), $m$ being incremented from top to bottom, and the second subscript $n$ refers to the period number in the $x$-direction. With these notations, the scattering process at the intersection between the $n^{th}$ hole chain and the $m^{th}$ MIM waveguide is described by four ingoing amplitudes, $A_{m,n}$, $B_{m+1,n}$, $C_{m,n+1}$, $D_{m,n}$, and four outgoing amplitudes, $A_{m+1,n}$, $B_{m,n}$, $C_{m,n}$ and $D_{m,n+1}$, which are coupled through five scattering coefficients, $\alpha$, $r_{sp}$, $t_{sp}$, $\rho$ and $\tau$, see Fig. 1 in the main text. The scattering coefficients are defined with a phase origin at the center of the scatterer, i.e., the intersection between a hole chain and a MIM waveguide.

The scattering process is described by a four-port scattering matrix

$$\begin{pmatrix} A_{m+1,n} \\ B_{m,n} \\ C_{m,n} \\ D_{m,n+1} \end{pmatrix} = \begin{pmatrix} \tau v & \rho v & \alpha u^{1/2} v^{1/2} & \alpha u^{1/2} v^{1/2} \\ \rho v & \tau v & \alpha u^{1/2} v^{1/2} & \alpha u^{1/2} v^{1/2} \\ \alpha u^{1/2} v^{1/2} & \alpha u^{1/2} v^{1/2} & t_{\mathrm{sp}} u & r_{\mathrm{sp}} u \\ \alpha u^{1/2} v^{1/2} & \alpha u^{1/2} v^{1/2} & r_{\mathrm{sp}} u & t_{\mathrm{sp}} u \end{pmatrix} \begin{pmatrix} A_{m,n} \\ B_{m+1,n} \\ C_{m,n+1} \\ D_{m,n} \end{pmatrix},$$

(1)

where $v = \exp(ik_0 n_h a_z)$, $u = \exp(ik_0 n_{\mathrm{sp}} a_x)$, $n_h$ is the normalized propagation constant of the TE$_{01}$ super-mode, $n_{\mathrm{sp}}$ is the normalized propagation constant of the gap-SPP and $k_0 = 2\pi/\lambda$ is the vacuum wavevector. In the 4x4 scattering matrix, the 2x2 top-left and bottom-right blocks represent the scattering along the longitudinal and transversal directions, respectively. The off-diagonal 2x2 blocks represent the coupling between both channels. Note that, to be valid, Eq. (1) does not require the structure to be periodic along $x$ and/or $z$-directions.

For the fishnet, because of the periodicity (pitch $a_x$) in the $x$-direction, the Floquet-Bloch condition imposes the pseudo-periodicity on the field amplitudes, and we have

$$C_{m,n+1} = \exp(ik_x a_x)\, C_{m,n} \quad \text{and} \quad D_{m,n+1} = \exp(ik_x a_x)\, D_{m,n},$$

(2)

with $k_x$ the $x$-component of the wavevector of the incident plane wave. Normal incidence ($k_x = 0$) will be assumed hereafter, but exactly the same derivation can be performed for $k_x \neq 0$. Reporting Eq. (2) into Eq. (1), one gets

$$C_{m,n} = D_{m,n} = \frac{\gamma v^{1/2}}{2\alpha u^{1/2}}\left(A_{m,n} + B_{m+1,n}\right),$$

(3)

and after elimination of the field amplitudes $C_{m,n}$ and $D_{m,n}$, the 4x4 scattering matrix reduces to a 2x2 scattering matrix, which only relates the field amplitudes in the longitudinal channel,

$$\begin{pmatrix} A_{m+1,n} \\ B_{m,n} \end{pmatrix} = \begin{pmatrix} (\tau + \gamma) v & (\rho + \gamma) v \\ (\rho + \gamma) v & (\tau + \gamma) v \end{pmatrix} \begin{pmatrix} A_{m,n} \\ B_{m+1,n} \end{pmatrix}.$$

(4)

In Eqs. (3) and (4), the new parameter $\gamma$ that results from the transversal coupling between the adjacent hole chains through the gap-SPP physically represents a multiple scattering term given by

$$\gamma = \frac{2\alpha^2 u}{1 - (t_{sp} + r_{sp})u}.$$

(5)

The fishnet effective index $n_{\mathrm{eff}}$ can be derived by applying Floquet-Bloch conditions in the longitudinal $z$-direction,

$$\begin{pmatrix} A_{m+1,n} \\ B_{m+1,n} \end{pmatrix} = e^{\pm ik_0 n_{\text{eff}} a_z} \begin{pmatrix} A_{m,n} \\ B_{m,n} \end{pmatrix}, \tag{6}$$

implying that $\exp(\pm ik_0 n_{\text{eff}} a_z)$ are the two eigenvalues of the transfer matrix that links the amplitudes of the $TE_{01}$ super-mode in two metallic layers separated by one period $a_z$. This transfer matrix is easily derived from the scattering matrix in Eq. (4) and reads as

$$\begin{pmatrix} A_{m+1,n} \\ B_{m+1,n} \end{pmatrix} = \frac{1}{(\tau + \gamma)v} \begin{pmatrix} [(\tau + \gamma)^2 - (\rho + \gamma)^2 \;]v^2 & (\rho + \gamma)v \\ -(\rho + \gamma)v & 1 \end{pmatrix} \begin{pmatrix} A_{m,n} \\ B_{m,n} \end{pmatrix}. \tag{7}$$

The trace of the matrix being equal to the sum of its eigenvalues, one straightforwardly obtains a closed-form expression for $n_{\text{eff}}$,

$$\cos(k_0 n_{\text{eff}} a_z) = \frac{(\tau + \gamma)^2 v^2 - (\rho + \gamma)^2 v^2 + 1}{2(\tau + \gamma)v}. \tag{8}$$

Further simple algebraic derivations lead to the calculation of the transfer-matrix eigenvectors, i.e., the fishnet Bloch modes. For the forward-propagating mode related to the $\exp(ik_0 n_{\text{eff}} a_z)$ eigenvalue, we obtain

$$\frac{B_{m,n}}{A_{m,n}} = \frac{(\rho + \gamma)v}{1 - (\tau + \gamma)v \exp(ik_0 n_{\text{eff}} a_z)}, \tag{9a}$$

$$\frac{C_{m,n}}{A_{m,n}} = \frac{\gamma v^{1/2}}{2\alpha u^{1/2}} \left( 1 + \frac{B_{m,n}}{A_{m,n}} e^{ik_0 n_{\text{eff}} a_z} \right). \tag{9b}$$

Equation (9b) has been used to plot the solid red curve in Fig. 3(b) that shows the resonance of the gap-SPP intensity in the structure.

## 2. Numerical accuracy and convergence

There is a slight difference (see Fig. 2 in the main text) between the model predictions for the fishnet effective index $n_{\text{eff}}$ and the fully-vectorial computational results obtained with the RCWA [2-4]. We mainly attribute the difference to the finite thickness of the metallic layers, which longitudinally couple the gap-SPP modes in the fishnet, an effect that is not taken into account in the model. Importantly we have checked that the difference does not result from an artifact due to numerical inaccuracies.

In our computations that rely on several original methods [2-8], numerical inaccuracies originate from the inevitable truncation of Fourier series that are used to expand the electromagnetic fields for numerical purposes. In order to evaluate the precision of the computational results, convergence tests have been performed by progressively increasing the number $N_x$ or $N_y$ of Fourier coefficients retained in the computation [Fourier series goes from $-(N-1)/2$ to $+(N-1)/2$]. Figure B shows the convergence curve of $\text{Re}(n_{\text{eff}})$ and $\text{Im}(n_{\text{eff}})$ calculated with the fully-vectorial approach for $\lambda = 2$ μm. For $N_x \times N_y = 71 \times 51$ Fourier harmonics, the relative error can be expected to be in the range of 1%, well smaller than the slight difference with the model predictions. We have checked that all computational results show similar numerical accuracy (not shown here).

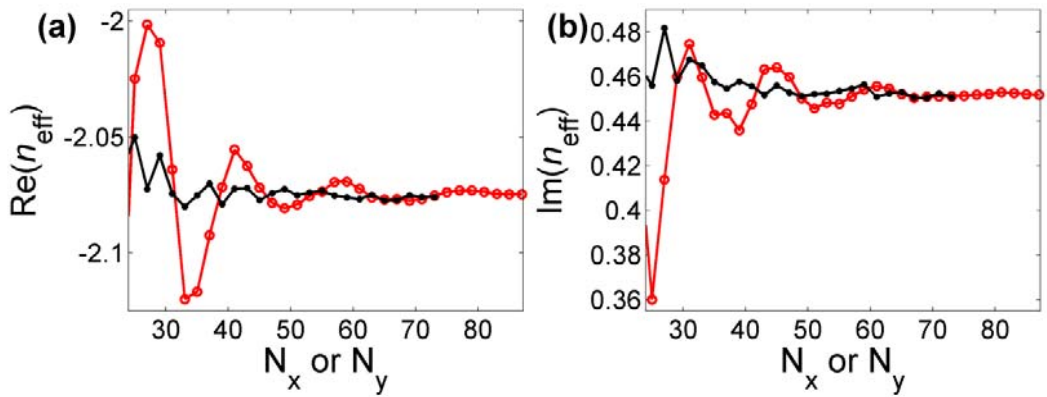

**Figure B.** Convergence tests for $Re(n_{eff})$ (a) and $Im(n_{eff})$ (b). The red circles show the convergence as a function of $N_x$ for $N_y = 41$ while the black dots show the convergence as a function of $N_y$ for $N_x = 61$.